\documentclass[aps,prl,twocolumn,amsmath,longbibliography]{revtex4-2}
\usepackage{graphicx}
\usepackage{subfigure}
\usepackage{url}
\usepackage{dcolumn}
\usepackage{bm}
\usepackage[]{natbib}
\usepackage{bbm}
\usepackage{mathrsfs}
\usepackage{amsmath}
\usepackage{amsfonts}
\usepackage[colorlinks=true, 
citecolor=blue, 
anchorcolor=blue, 
linkcolor=blue, 
urlcolor=blue]{hyperref}
\usepackage{graphicx,epstopdf}
\usepackage{subfigure}
\usepackage{epsfig}
\usepackage{bm}
\usepackage{color}
\usepackage{natbib}
\usepackage{amssymb}
\usepackage{xcolor}
\usepackage{braket}
\usepackage{tabularx}
\usepackage{multirow}
\usepackage{booktabs}
\usepackage{makecell}
\newcommand{\CenterCell}[1]{%
  \vspace{0pt}%
  \begin{tabular}{@{}c@{}}%
    #1
  \end{tabular}%
  \vspace{0pt}%
}
\begin{document}

\title{Alteraxial Phonons in Collinear Magnets}

\author{Fuyi Wang$^{1,\dagger}$, Junqi Xu$^{1,\dagger}$, Xinqi Liu$^{1}$, Lifa Zhang$^{2}$, Huaiqiang Wang$^{2,4\ast}$ and Haijun Zhang$^{1,3,4,5,\ast}$}

\affiliation{
$^1$ National Laboratory of Solid State Microstructures, School of Physics, Nanjing University, Nanjing 210093, China\\
$^2$ Center for Quantum Transport and Thermal Energy Science, Institute of Physics Frontiers and Interdisciplinary Sciences, School of Physics and Technology, Nanjing Normal University, Nanjing 210023, China\\
$^3$ Collaborative Innovation Center of Advanced Microstructures, Nanjing University, Nanjing 210093, China\\
$^4$ Jiangsu Physical Science Research Center, Nanjing 210093, China\\
$^5$ Jiangsu Key Laboratory of Quantum Information Science and Technology, Nanjing University, China}

\email{zhanghj@nju.edu.cn}
\email{hqwang@njnu.edu.cn}


\begin{abstract}
Axial phonons, carrying angular momentum through rotational lattice vibrations, offer a promising platform for exploring phonon-magnetic coupling effects. However, how the interplay of lattice and magnetism determine the phonon angular momentum (PAM) of axial phonons remains elusive. Here, based on magnetic point group theory, we establish a symmetry framework to classify phonons in collinear magnets (e.g. ferromagnets, antiferromangets, altermagnets) into three distinct categories: ferroaxial, antiferro-nonaxial, and alteraxial phonons, which are distinguished by their different PAM patterns. Beyond the ferroaxial phonons featuring $s$-wave PAM, we reveal a complete series of alteraxial phonons, characterized by higher-order-wave PAM patterns ranging from $p$- to $j$-wave. Notably, alteraxial phonons are not limited to altermagnets, but also emerge in ferromagnets and antiferromagets. Our high-throughput search predicts hundreds of candidate magnetic materials hosting alteraxial phonons. \emph{Ab initio} calculations on representative magnets further confirm the existence and distinct symmetry-enforced nodal structures of PAM in alteraxial phonons. Our work provides a complete classification for axial phonons in collinear magnetic systems and paves the way for engineering magneto-phononic phenomena.
\end{abstract}
\maketitle

\noindent{\bf{Introduction.}} Phonons with nonzero angular momentum stemming from rotational lattice vabirations~\cite{zhang_angular_2014,zhang_chiral_2015,zhu_observation_2018}, recently termed axial phonons~\cite{juraschek_chiral_2025}, have attracted growing research interest over the past decade~\cite{chen_entanglement_2019,grissonnanche_chiral_2020,ishito_truly_2023,shabala_phonon_2024,kahana_light_2024,klebl_ultrafast_2025,paiva_dynamically_2025,ueda_chiral_2023, hernandez_chiral_2023,kim_chiral-phonon-activated_2023,luo_transverse_2023,hernandez_observation_2023,romao_phonon-induced_2024,basini_terahertz_2024,paiva_dynamically_2025,klebl_ultrafast_2025,yang_catalogue_2025,chen_emergence_2025,zhang_advances_2025,suzuki_disorder_2025}. Intriguingly, phonon angular momentum (PAM) can yield an effective phonon magnetic moment\cite{luo_large_2023,hernandez_chiral_2023,ren_phonon_2021,zhang_gate-tunable_2023,tang_exciton_2024,mustafa_origin_2025,xue_extrinsic_2025}, which enables the coupling of axial phonons to both external magnetic fields and internal magnetic moments~\cite{park_phonon_2020,yin_chiral_2021,flebus_phonon_2023,bonini_frequency_2023,davies_phononic_2024,tauchert_polarized_2022,lujan_spinorbit_2024,shabala_phonon_2024,ren_adiabatic_2024,bowen_chiral_2024,okamoto_altermagnetic_2025,shabala_axial_2025,zhang_comprehensive_2025}. Such couplings have been shown to give rise to rich phono-magnetic effects, including the phonon Zeeman effect~\cite{juraschek_giant_2022, wang_ab_2025}, the ultrafast Einstein-de Haas effect~\cite{dornes_ultrafast_2019,tauchert_polarized_2022} and the phonon Barnett effect~\cite{luo_large_2023,davies_phononic_2024}. Specifically, the phonon Zeeman effect can induce observable frequency splittings between left- and right-circularly polarized phonons in magnetic materials, as demonstrated in ferromagnetic Weyl semimetal Co$_3$Sn$_2$S$_2$~\cite{che_magnetic_2025,yang_inherent_2025,zhang_general_2025}. Despite recent progress, the interplay between axial phonons and magnetic order still remains largely unexplored. 

\begin{figure}[t]
	\centering
	\includegraphics[width=0.47\textwidth]{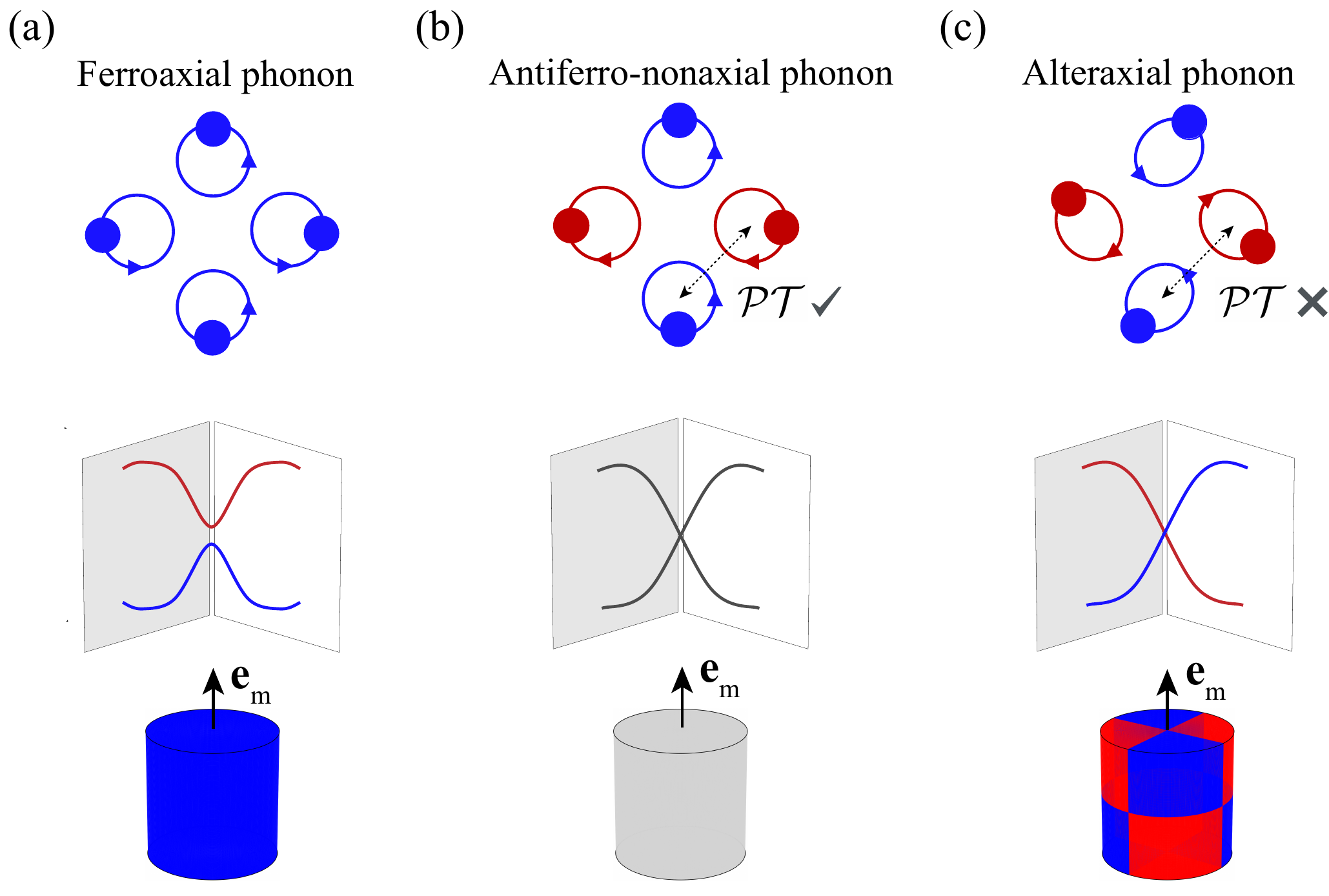}
	\caption{Schematics of three distinct types of phonons by PAM configurations in collinear magnets. Illustration of the lattice vibrations (top row), $\Gamma$-centered phonon spectra (middle row) and corresponding phonon angular momentum (PAM) patterns in reciprocal space (bottom row) for (a) ferroaxial, (b) antiferro-nonaxial and (c) alteraxial phonons, respectively. The magnetization direction ($\bf{e}_\mathrm{m}$) is set as the quantization axis for the PAM. Blue and red colors represent the left- and right-handed circular polarization with positive and negative PAM, respectively.}\label{fig1}
\end{figure}
\begin{table*}[t]
\renewcommand{\arraystretch}{2} 
\caption{Complete classification of phonons in collinear magnets based on magnetic point group symmetry. According to the symmetry conditions and the PAM wave pattern (even or odd), phonons in collinear magnets are categorized into four types: (i) $\mathcal{PT}$-symmetric nonaxial phonon; (ii) Centrosymmetric (CS) even-wave axial phonon; (iii) Noncentrosymmetric (NCS) odd-wave axial phonon; (iv) NCS even-wave axial phonon. The nonaxial PAM pattern corresponds to antiferro-nonaxial phonons. The $s$-wave patterns describe ferroaxial phonons, whereas all higher-order-wave patterns ($p$-wave to $j$-wave) correspond to alteraxial phonons. Magnetic point groups (MPGs) are given in the UNI notation~\cite{campbell_introducing_2022}. Parentheses are used to specify the orientation of the corresponding magnetic axis.} \label{tab:wave_type}
\centering
\begin{tabular}{c|c|c|c}
\hline
Phonon type & Wave & Pattern & Magnetic point groups\\
\hline
\makecell{$\mathcal{PT}$-symmetric\\ nonaxial phonon} & N/A & \raisebox{-1.5ex}{\includegraphics[height=0.6cm]{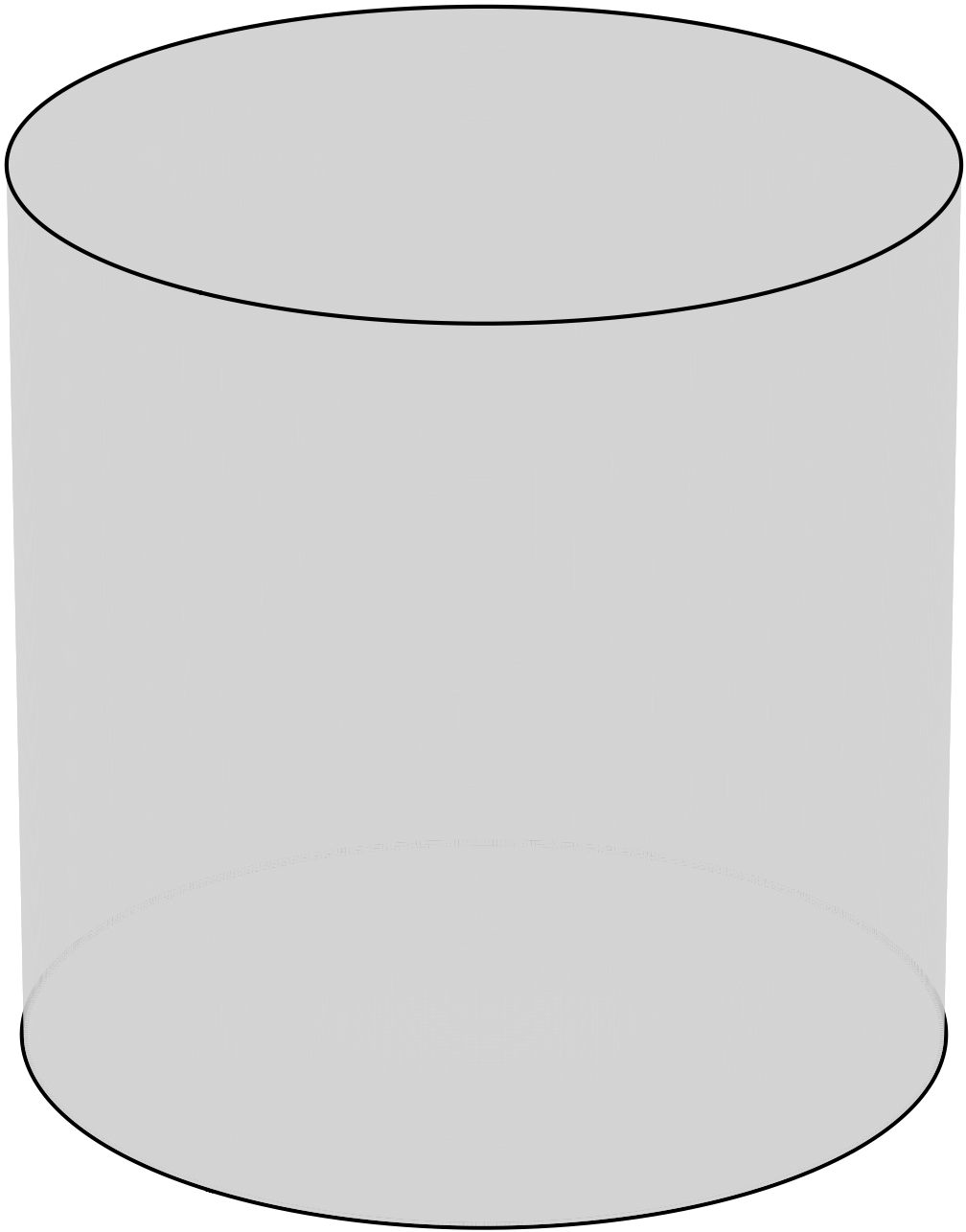}} & \gape{\makecell{4/m.1$'$, $\overline{3}$m.1$'$, 6/m$'$m$'$m$'$, 4/m$'$, 6/m.1$'$, 6/m$'$, mmm.1$'$, 2/m.1$'$, 6$'$/m, m$'$m$'$m$'$, 4$'$/m$'$,\\ 4$'$/m$'$m$'$m, 4/mmm.1$'$, 4/m$'$mm, 6/m$'$mm, 2/m$'$, 2$'$/m, m$'$mm, $\overline{3}$$'$m, $\overline{3}$.1$'$, $\overline{3}$$'$m$'$, $\overline{3}$$'$, $\overline{1}$.1$'$, \\4/m$'$m$'$m$'$, $\overline{1}$$'$, 6/mmm.1$'$, 6$'$/mmm$'$}}\\
\hline
\raisebox{3ex}{\multirow{6}{*}{\makecell{CS even-wave\\ axial phonon}}} & \CenterCell{s} & \raisebox{-1.5ex}{\includegraphics[height=0.6cm]{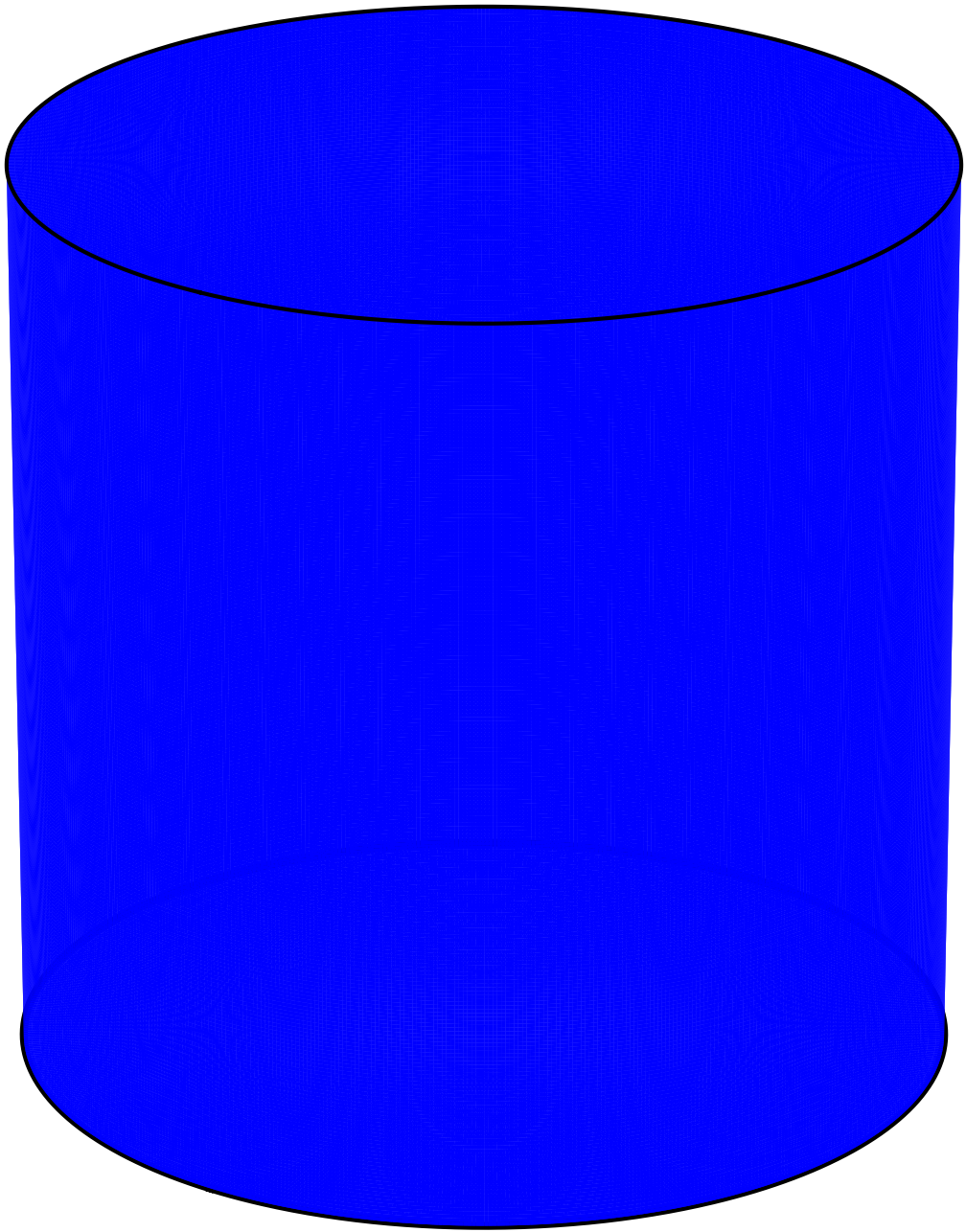}} & \CenterCell{2$'$/m$'$(xz), 6/mm$'$m$'$, 4/mm$'$m$'$, 4/m.1, $\overline{3}$.1, $\overline{1}$.1, m$'$m$'$m(z), 6/m.1, 2/m.1(y), $\overline{3}$m$'$} \\
\cline{2-4}
& \multirow{1}{*}{d} & \raisebox{-1.5ex}{\includegraphics[height=0.6cm]{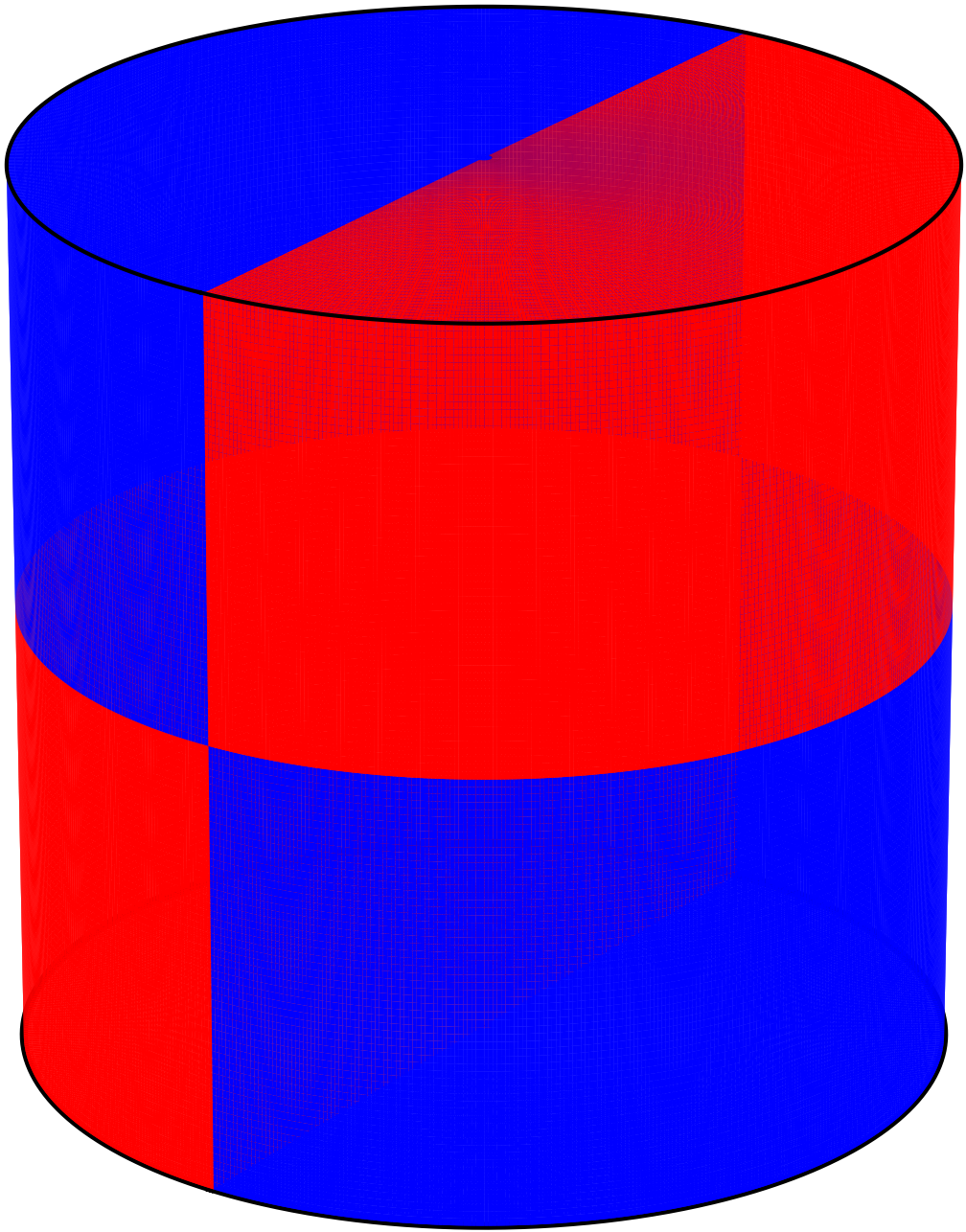} \includegraphics[height=0.6cm]{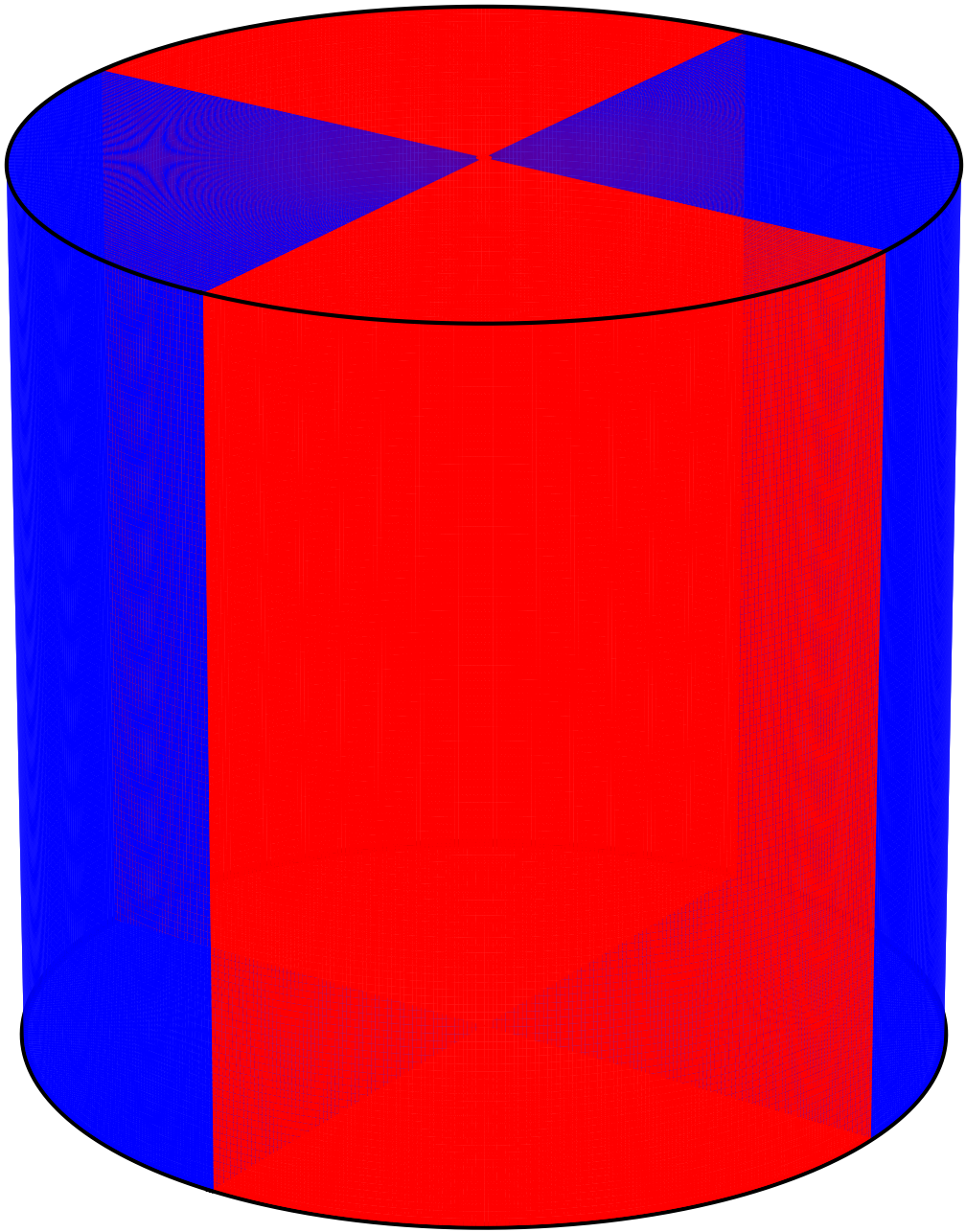}} & \CenterCell{mmm.1(xy), m$'$m$'$m(xy), 2/m.1(xz), 2$'$/m$'$(y), mmm.1(z), 4$'$/m, 4$'$/mm$'$m} \\
\cline{2-4}
& \multirow{1}{*}{g} & \raisebox{-1.5ex}{\includegraphics[height=0.6cm]{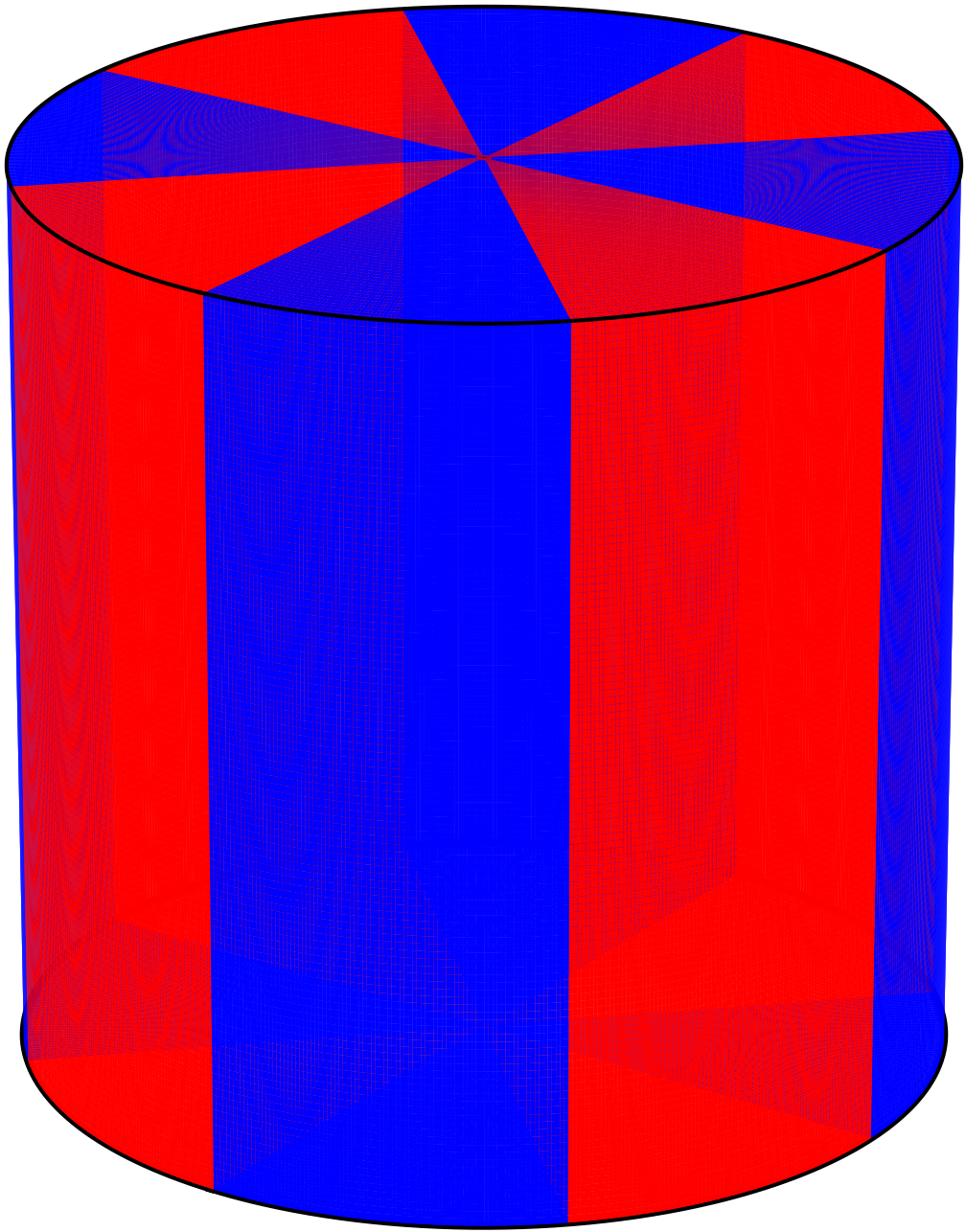} \includegraphics[height=0.6cm]{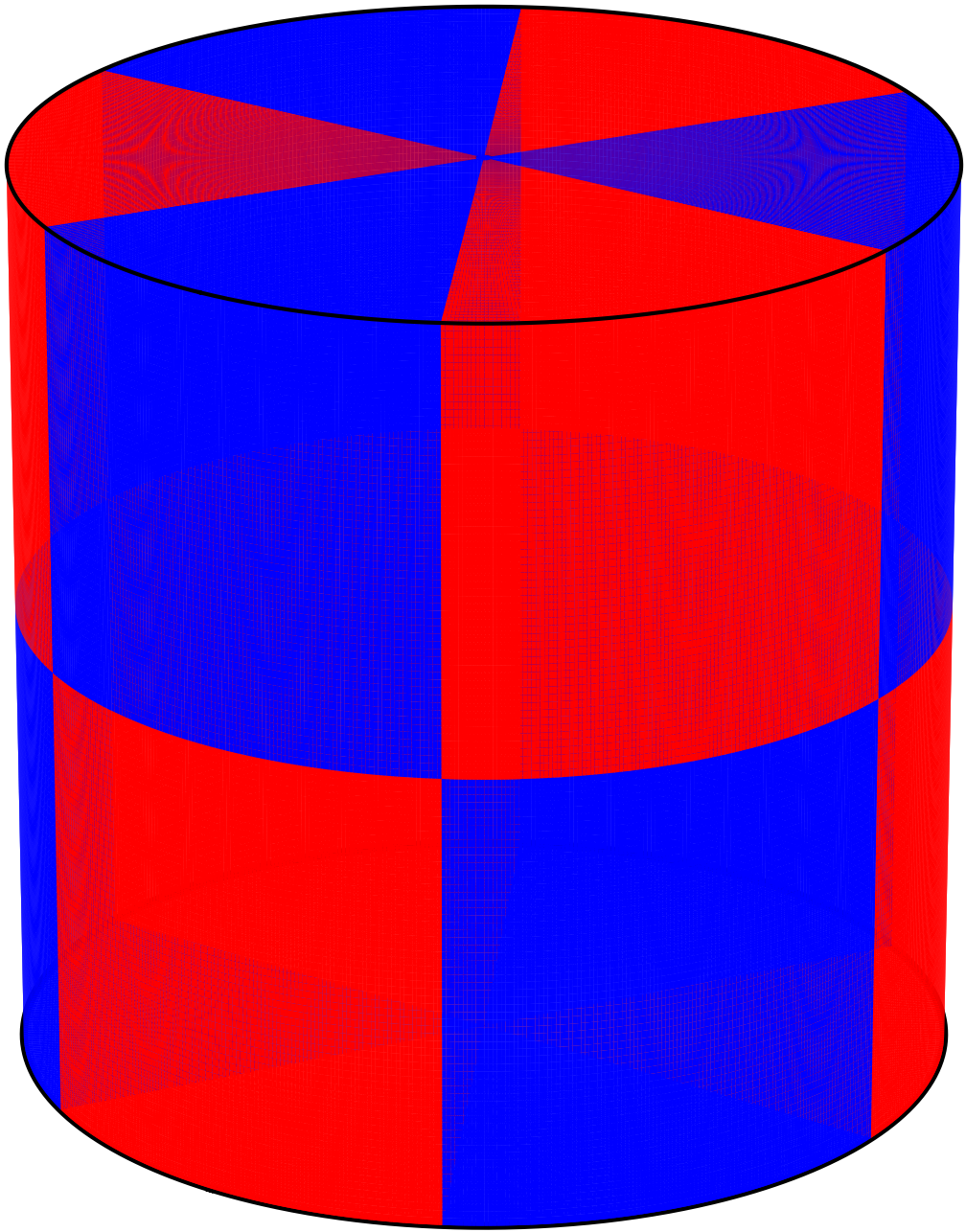}} & \CenterCell{4/mmm.1, $\overline{3}$m.1, 6$'$/m$'$mm$'$, 6$'$/m$'$} \\
\cline{2-4}
& \CenterCell{i} & \raisebox{-1.5ex}{\includegraphics[height=0.6cm]{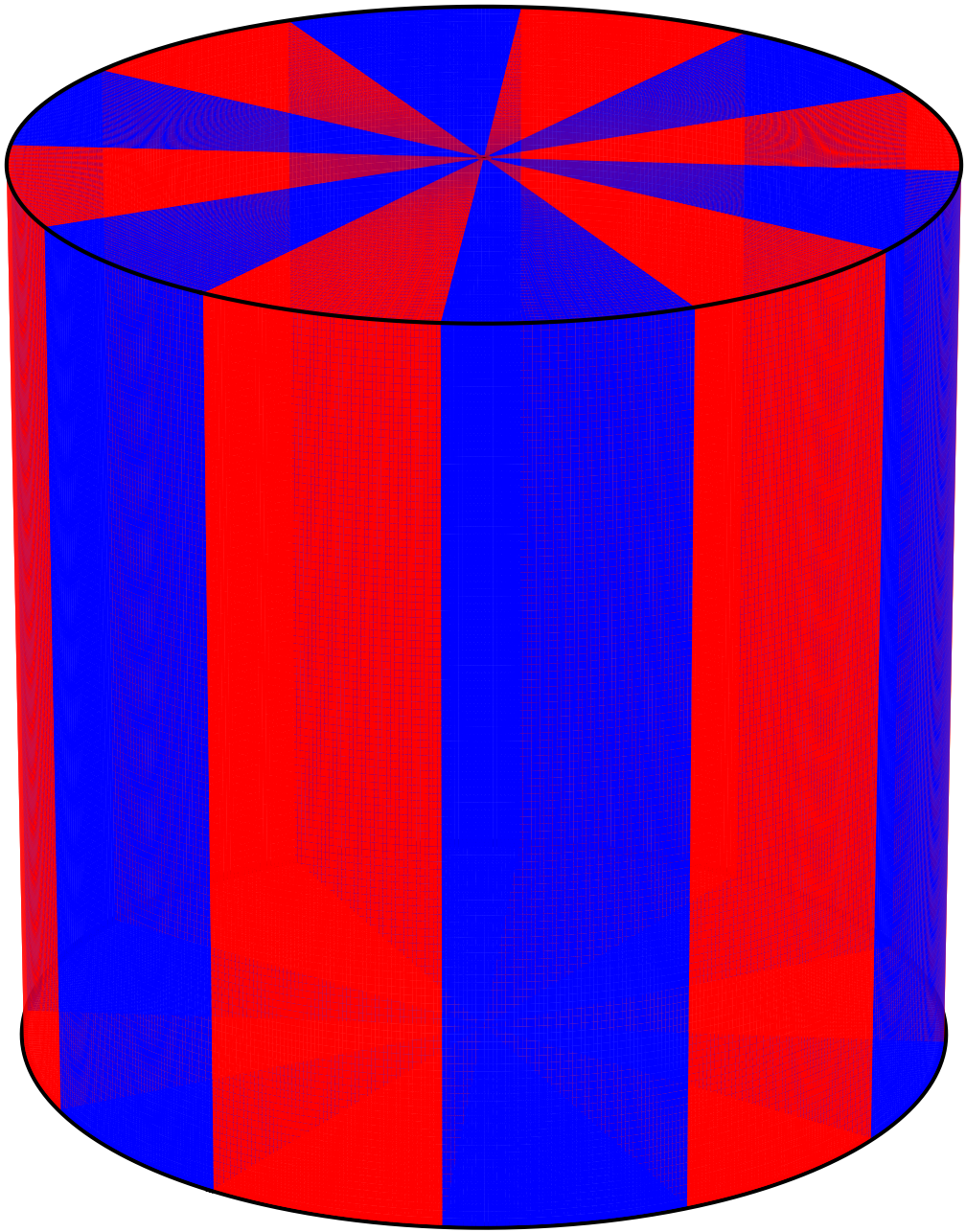}} & \CenterCell{6/mmm.1} \\
\hline
\raisebox{3ex}{\multirow{6}{*}{\makecell{NCS odd-wave\\ axial phonon}}} & \CenterCell{p} & \raisebox{-1.5ex}{\includegraphics[height=0.6cm]{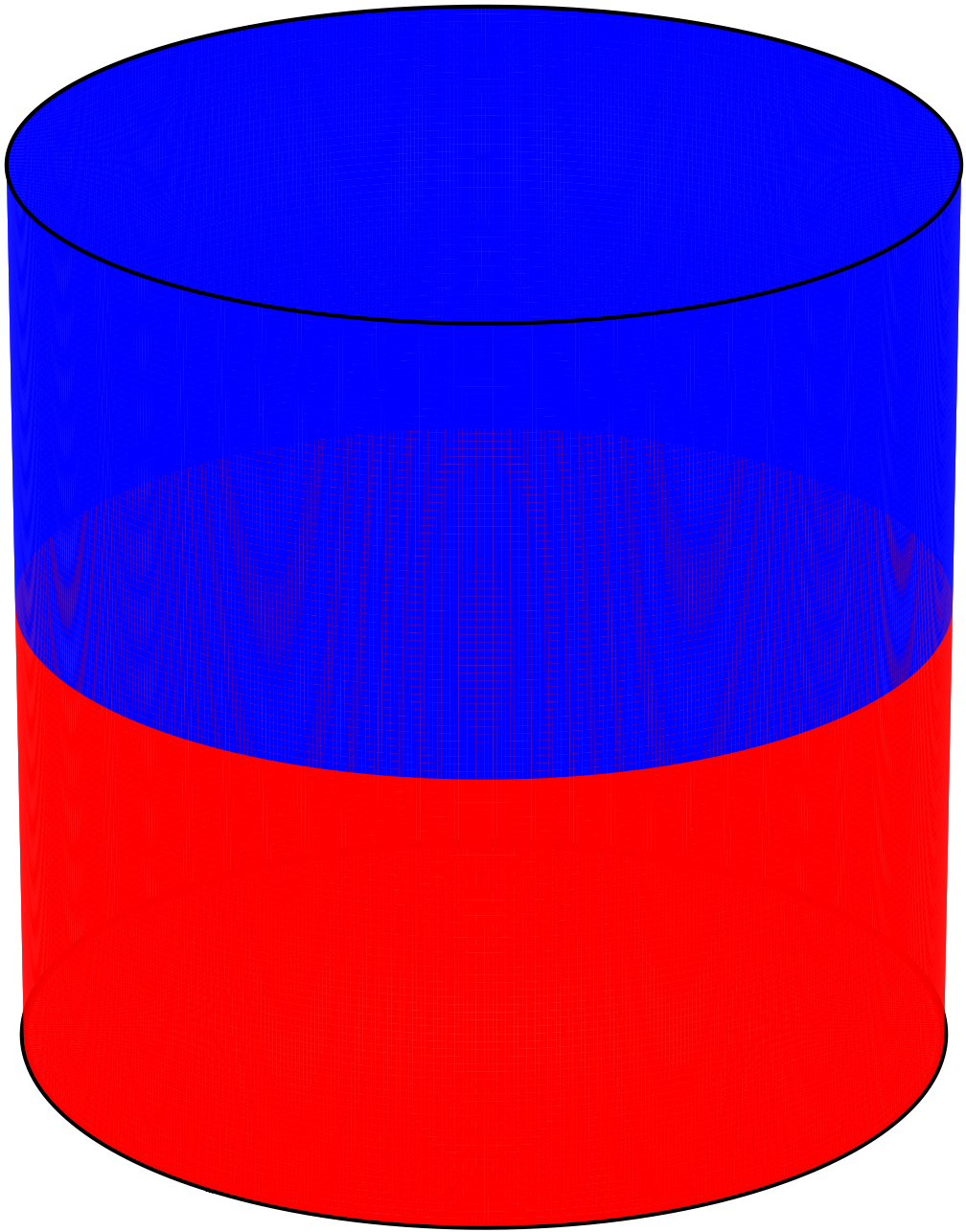} \includegraphics[height=0.6cm]{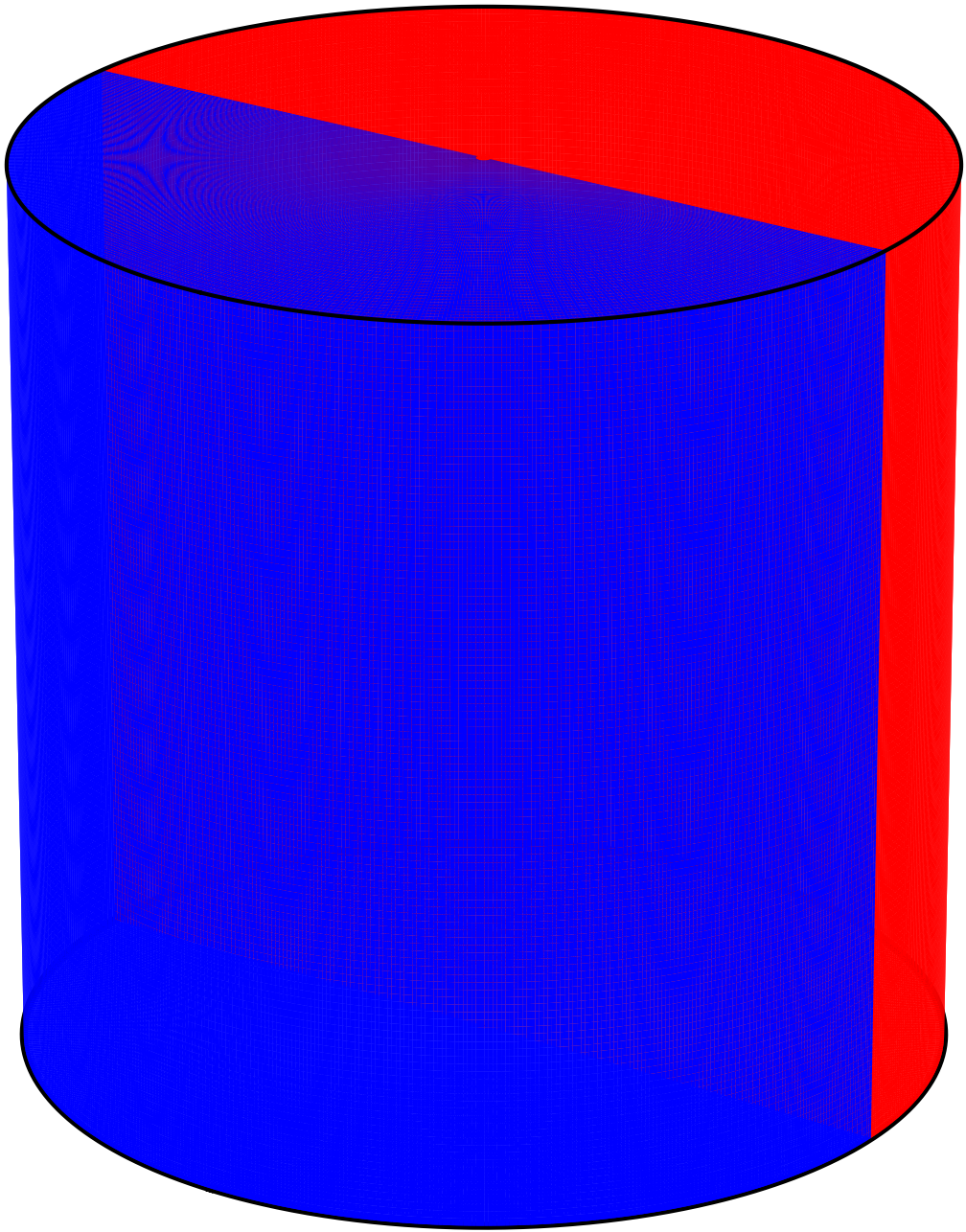}} & \gape{\makecell{4$'$22$'$, 32.1, 32.1$'$, 422.1, 6.1$'$, 622.1, 6$'$, 3.1$'$, 422.1$'$, 2.1$'$, 622.1$'$, m.1$'$, 222.1$'$, 4$'$, 1.1$'$,\\ 4.1$'$, 2.1(xz), 6$'$22$'$, 222.1, m$'$m2$'$(x), 222.1$'$, mm2.1(xy), m$'$m$'$2(xy), m$'$(y), \\2$'$2$'$2(xy), 2.1$'$, 2$'$(y), m.1(xz), mm2.1$'$(xy), m.1$'$, 222.1, 222.1$'$, 222.1}} \\
\cline{2-4}
& \multirow{1}{*}{f} & \raisebox{-1.5ex}{\includegraphics[height=0.6cm]{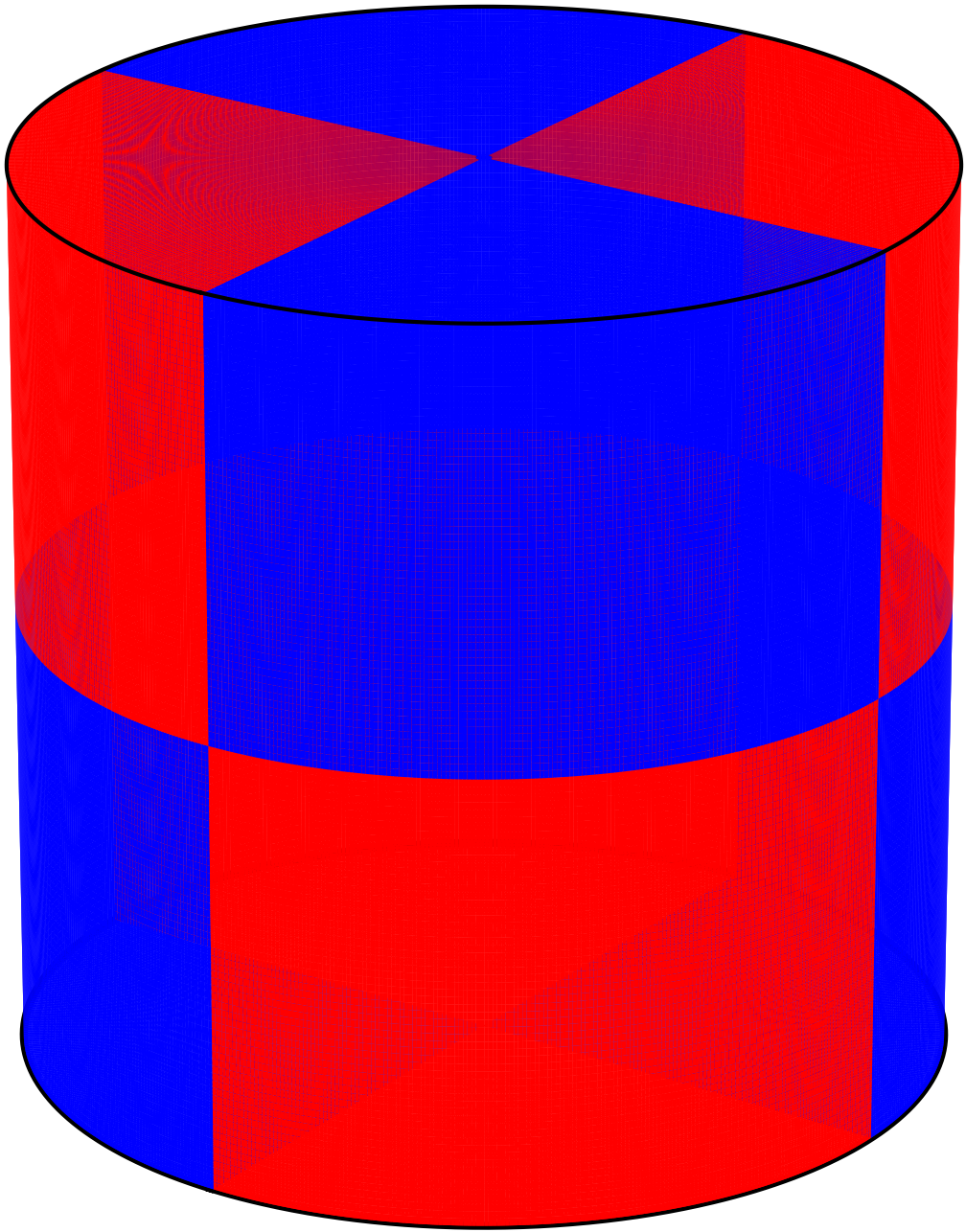} \includegraphics[height=0.6cm]{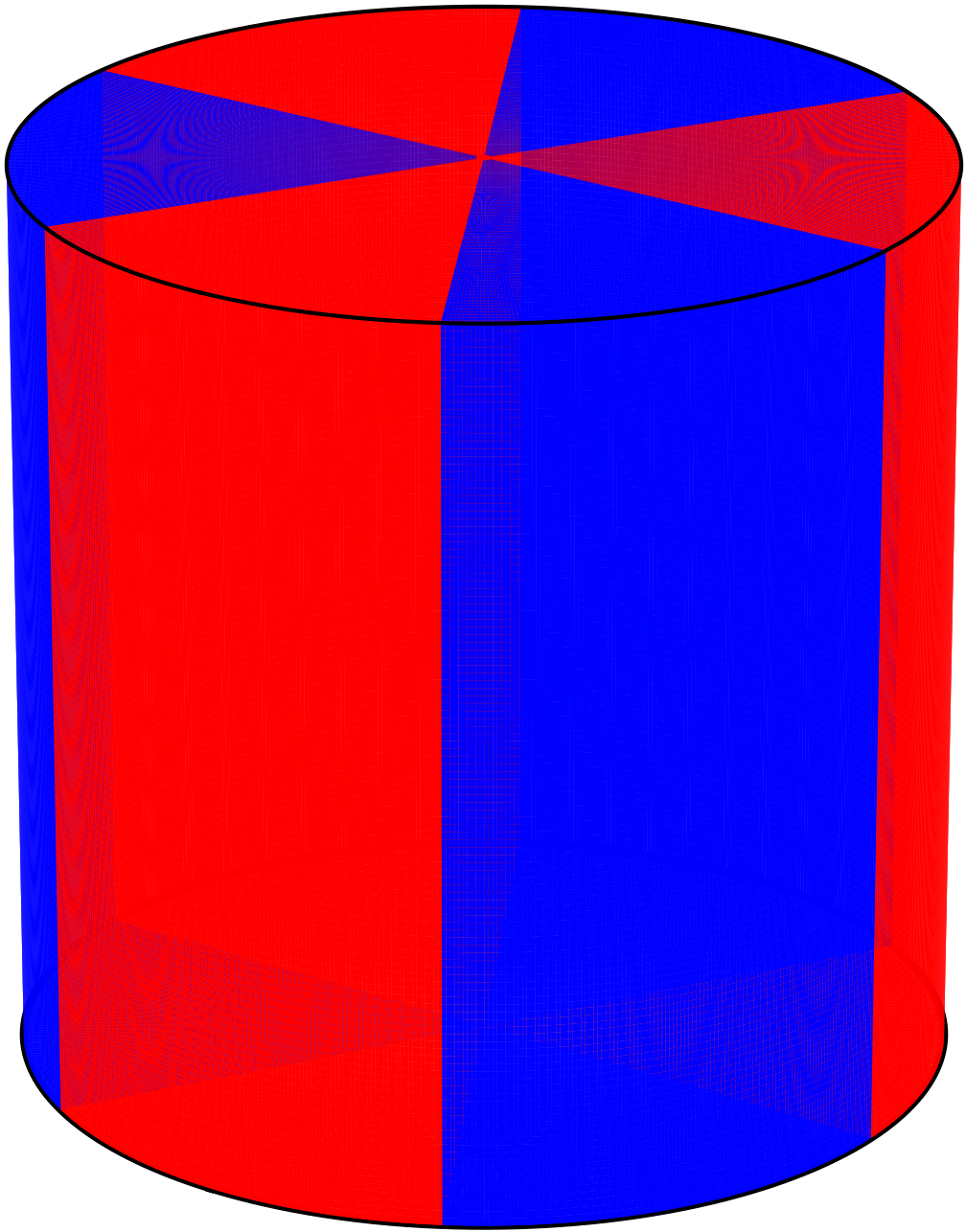}} & \CenterCell{mm2.1$'$(z), $\overline{4}$.1$'$, $\overline{4}$2m.1$'$, $\overline{4}$2m.1,3m.1$'$, $\overline{6}$m2.1$'$, 3m.1, $\overline{6}$m2.1, $\overline{6}$$'$m2$'$, $\overline{6}$$'$m$'$2, $\overline{6}$.1$'$, $\overline{6}$$'$} \\
\cline{2-4}
& \CenterCell{h} & \raisebox{-1.5ex}{\includegraphics[height=0.6cm]{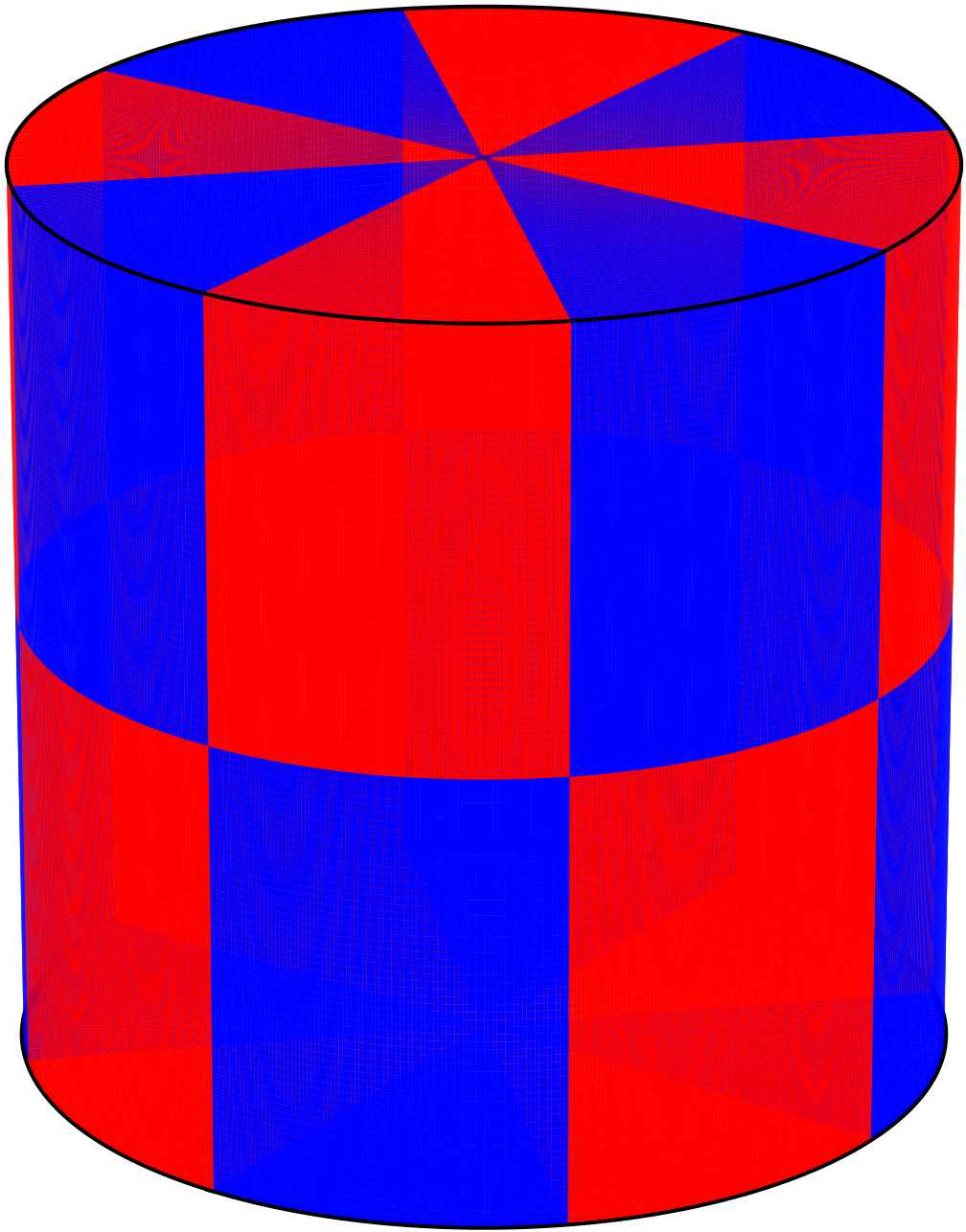}} & \CenterCell{4mm.1$'$} \\
\cline{2-4}
& \CenterCell{j} & \raisebox{-1.5ex}{\includegraphics[height=0.6cm]{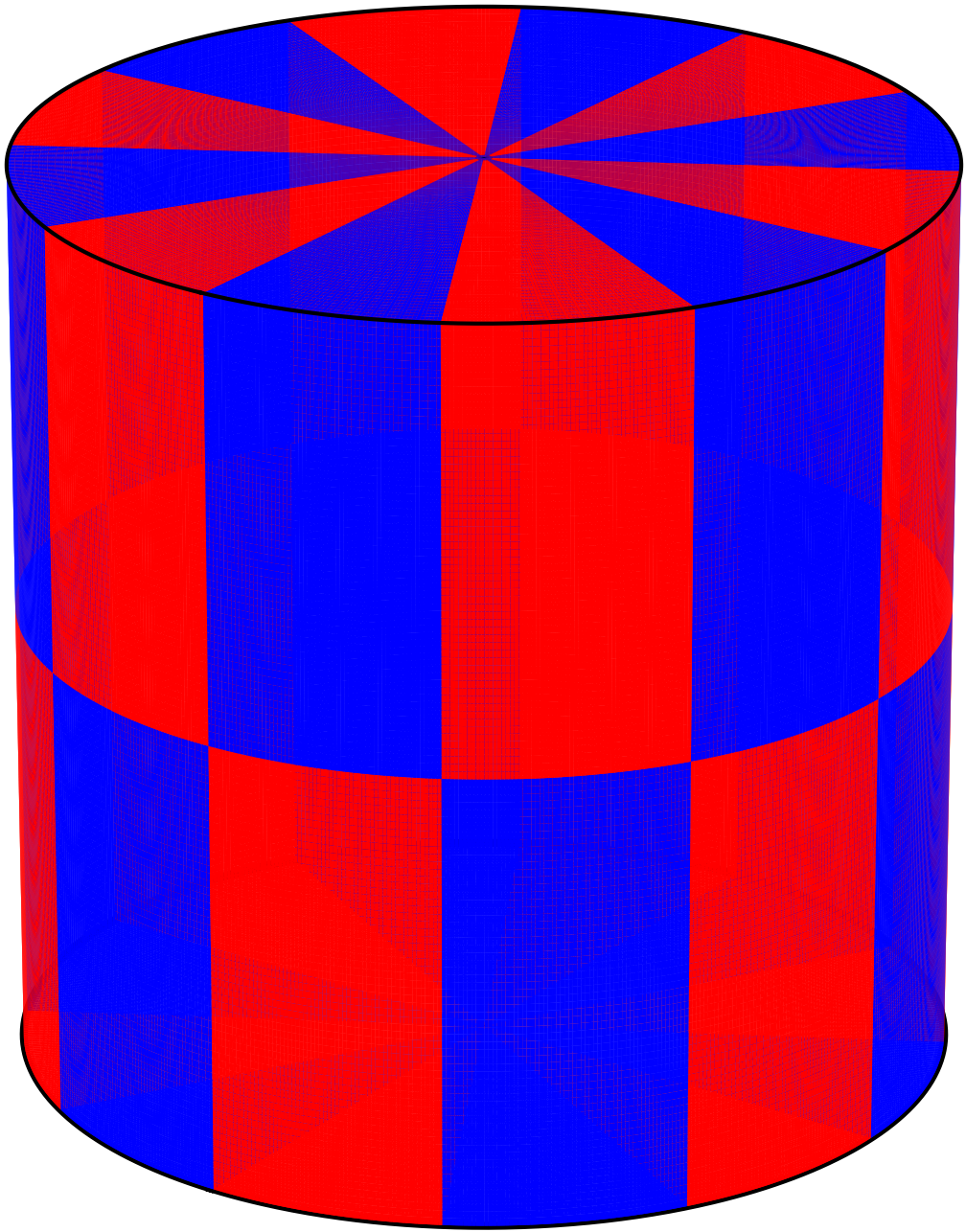}} & \CenterCell{6mm.1$'$} \\
\hline
\raisebox{3ex}{\multirow{6}{*}{\makecell{NCS even-wave\\ axial phonon}}} & \CenterCell{s} & \raisebox{-1.5ex}{\includegraphics[height=0.6cm]{s.png}} & \makecell{m.1(y), $\overline{6}$.1, 2$'$(xz), m$'$m$'$2(z), $\overline{6}$m$'$2$'$, 3.1, 2$'$2$'$2(z), 4.1, m$'$(xz), $\overline{42}$$'$m$'$,\\ 62$'$2$'$, 6m$'$m$'$, 4m$'$m$'$, 2.1(y), 42$'$2$'$, m$'$m2$'$(y), 32$'$, $\overline{4}$.1, 3m$'$, 1.1, 6.1} \\
\cline{2-4}
& \multirow{1}{*}{d} & \raisebox{-1.5ex}{\includegraphics[height=0.6cm]{d_x_y.png} \includegraphics[height=0.6cm]{d_y_z.png}} & \CenterCell{$\overline{4}$$'$2m$'$, mm2.1(z), $\overline{4}$$'$, 4$'$m$'$m, $\overline{4}$$'$2$'$m, m$'$m2$'$(z)} \\
\cline{2-4}
& \multirow{1}{*}{g} & \raisebox{-1.5ex}{\includegraphics[height=0.6cm]{g_x_y__-x__2_+_y__2_.png} \includegraphics[height=0.6cm]{g_x_z__-x__2_+_3_y__2_.png}} & \CenterCell{4mm.1,6$'$mm$'$} \\
\cline{2-4}
& \CenterCell{i} & \raisebox{-1.5ex}{\includegraphics[height=0.6cm]{i_x_y__3_x__4_-_10_x__2_y__2_+_3_y__4_.png}} & \CenterCell{6mm.1} \\
\hline
\end{tabular}
\end{table*}

A central open question in the study of axial phonons in magnets is how the lattice structure and magnetic order influence PAM. Given that both PAM and electron spin are axial vectors, they are expected to exhibit similar behaviors despite their essential differences. Collinear magnets are typically classified into three types based on their distinct spin orders, namely ferro-, antiferro-, and alter-magnets, which induce corresponding spin polarizations in their electronic band structures~\cite{smejkal_emerging_2022,smejkal_beyond_2022,song_altermagnets_2025}. In an analogous manner, phonons in collinear magnets can also be categorized into three types, namely, ferroaxial, antiferro-nonaxial and alteraxial phonons, distinguished by their PAM configurations and resulting phononic circular polarizations in the phonon spectra, as illustrated in Fig.~1. Ferroaxial phonons [Fig.~1(a)] are characterized by rotational lattice vibrations of a single handedness, yielding circularly polarized phonon spectra hosting $s$-wave PAM with a single sign. In contrast, antiferro-nonaxial phonons [Fig.~1(b)] feature $\mathcal{PT}$-symmetric rotational lattice vibrations which result in unpolarized phonon spectra with vanishing PAM pattern in reciprocal space. Notably, for rotational lattice vibrations with an alternating handedness that are not related by $\mathcal{PT}$ symmetry, the resulting phonon spectra exhibit circular polarizations with alternating signs and higher-order-wave PAM patterns [Fig.~1(c)]. We term this third category \emph{alteraxial phonons}. To our knowledge, a comprehensive group-based definition and study of axial phonons in magnets is still lacking, thereby significantly hindering the identification of magnetic materials hosting alteraxial phonons. 

\begin{figure*}[t]
	\centering
	\includegraphics[width=1.0\textwidth]{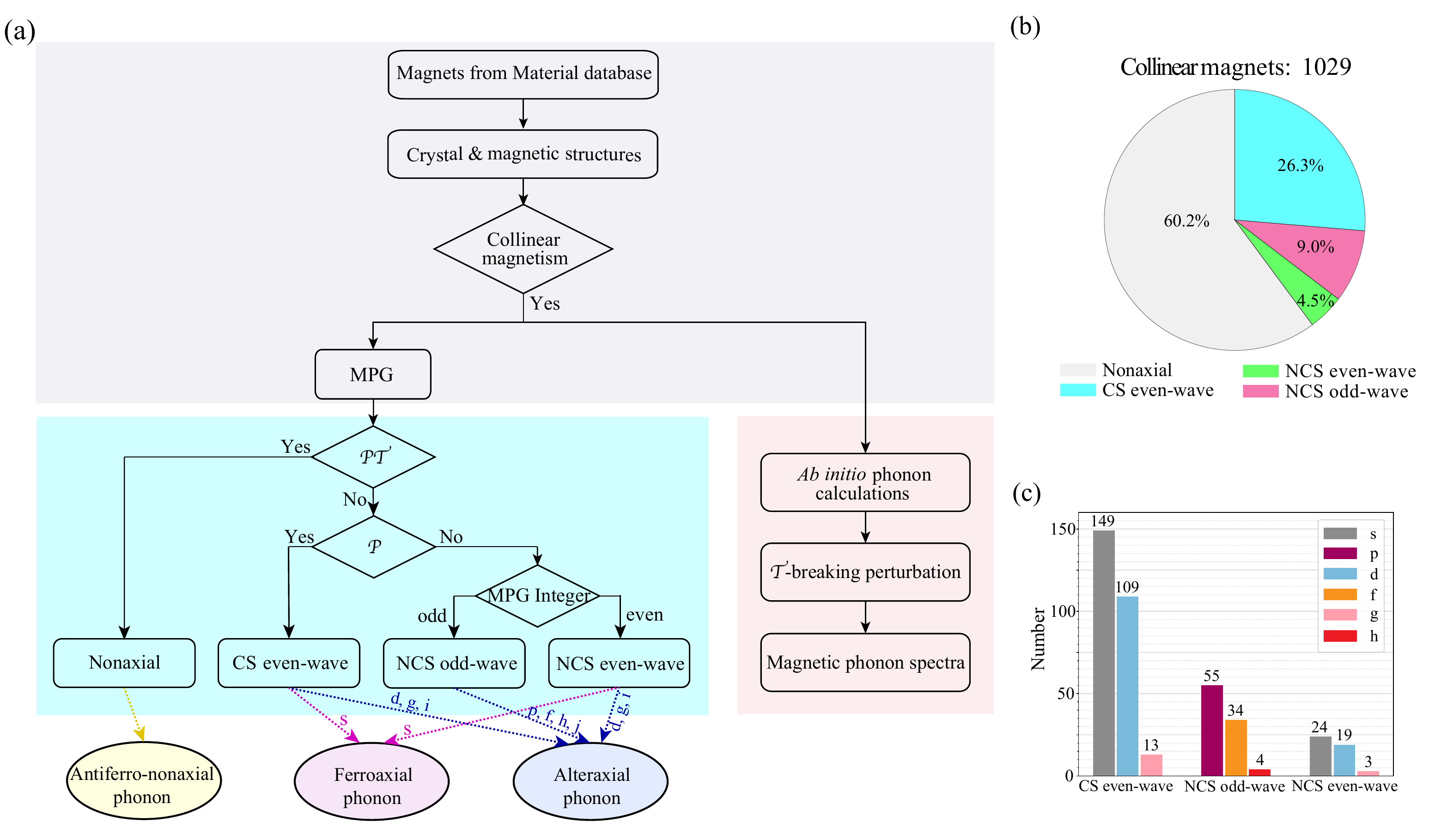}
	\caption{High-throughput phonon calculations in collinear magnets. The workflow (a) consists of three parts: The gray-shaded part filters out collinear magnets from material databases and provides their crystal and magnetic structures along with the corresponding magnetic point group (MPG); The cyan-shaded part classifies the phonon type according to the MPG, where the MPG integer denotes the number of symmetry-enforced zero-PAM nodal surfaces crossing the $\Gamma$ point; The pink-shaded part performs \emph{ab initio} calculations of magnetic phonon spectra. (b, c) High-throughput screening for all 1029 collinear magnets from the MAGNDATA database, with (b) distribution of materials across the four distinct phonon types and (c) number of materials hosting different wave patterns within the three axial phonon families.}\label{fig2}
\end{figure*}
In this work, based on magnetic point group (MPG) theory, we propose a systematic symmetry analysis of phonons in collinear magnets. Remarkably, beyond the ferroaxial ($s$-wave) phonon, we reveal a complete series of alteraxial phonons exhibiting higher-order-wave PAM, ranging from $p$- to $j$-wave. We have also performed high-throughput calculations and predicted hundreds of magnetic materials hosting alteraxial phonons.  \emph{Ab initio} magnetic phonon spectra and corresponding PAM patterns are further presented for representative materials, including CrSb, Cr$_2$SbAs, and MnSe. By establishing a systematic framework for axial phonons in collinear magnets, our work should open an avenue for studying new phonon-magnetic effects.\\

\noindent{\bf{MPG-based symmetry analysis.}} In collinear magnets, we define the magnetization direction (labeled as $\bf{e}_\mathrm{m}$  direction) as the quantization axis of the PAM. For a single phonon mode $\epsilon_{\bf{q}}$ with $\bf{q}$ denoting its wavevector, the magnitude of PAM corresponds to the phononic circular polarization along $\bf{e}_\mathrm{m}$ direction~\cite{zhang_chiral_2015}, which is given by
\begin{equation}
	j_{\textrm{m}}(\bf{q})=\langle \epsilon_{\bf{q}}|\ \big(|\textrm{R}_{\textrm{m}}\rangle\langle \textrm{R}_{\textrm{m}}|-| \textrm{L}_{\textrm{m}}\rangle\langle \textrm{L}_{\textrm{m}}|\big) \ |\epsilon_{\bf{q}}\rangle\hbar.
\end{equation}
Here, $|\epsilon_{\bf{q}}\rangle$ denotes the phonon-mode eigenvector, $\hbar$ is the Planck's constant, and $|\textrm{R}_{\textrm{m}}\rangle$ ($|\textrm{L}_{\textrm{m}}\rangle$) represents the right-handed (left-handed) circularly polarized eigenstate with angular momentum $\hbar$ ($-\hbar$) along $\bf{e}_\mathrm{m}$ direction. Throughout this work, we consider $\Gamma$-centered phonon modes, whose PAM symmetry properties are characterized by a MPG denoted as $G$~\cite{campbell_introducing_2022}. Note that our work differs from previous studies of chiral phonons in nonmagnetic systems~\cite{yang_catalogue_2025}, which focused primarily on their helicity from a symmetry perspective. Under a symmetry operation $\mathcal {G}\in G$, the PAM $j_{\mathrm{m}}(\bf{q})$ satisfies
\begin{equation}
j_{\textrm{m}}(\mathcal{G}\textbf{q})=\eta(\mathcal{T})\chi(\mathcal {R}) j_{\textrm{m}}(\textbf{q}) .
\end{equation}
Here, $\mathcal {T}$ is the time-reversal operation, and $\eta(\mathcal{T})=-1$ (+1) when  $\mathcal {T}$ is (not) contained in $\mathcal{G}$. $\mathcal {R}$ denotes the point-group part operation in $\mathcal{G}$, and $\chi(\mathcal{R})=+1$  $(-1)$ if $\mathcal{R}$ preserves (switches) the handedness of the circularly polarized states $|\textrm{R}_{\textrm{m}}\rangle$ and $|\textrm{L}_{\textrm{m}}\rangle$. Specifically, time-reversal symmetry $\mathcal{T}$ yields $j_{\textrm{m}}(-\textbf{q})=-j_{\textrm{m}}(\textbf{q})$, while inversion symmetry $\mathcal{P}$ gives $j_{\textrm{m}}(-\textbf{q})=j_{\textrm{m}}(\textbf{q})$. Consequently, in the presence of the combined $\mathcal{PT}$ symmetry, the PAM $j_{\textrm{m}}(\textbf{q})$ is forced to vanish, which corresponds to antiferro-nonaxial phonons in collinear magnets. 

Breaking the combined $\mathcal{PT}$ symmetry allows for a nonzero $j_{\textrm{m}}(\textbf{q})$, thereby leading to axial phonons encompassing both ferroaxial and alteraxial phonons. Interestingly, if at least one symmetry operation in $G$ yields $\eta(\mathcal{T})\chi(\mathcal{R}) = -1$, then $j_{\textrm{m}}(\textbf{q})$ should vanish at the $\Gamma$ point, while alternating in sign across the surrounding Brillouin zone (BZ).  This inevitably leads to nodal surfaces of vanishing $j_{\textrm{m}}(\textbf{q})$ passing through the $\Gamma$ point in the three-dimensional BZ [see the supplemental material (SM) for details~\cite{supp}]. The number of such nodal surfaces, termed an MPG integer determined from MPG constraints in Eq. (2), ranges from 1 to 7, defining alteraxial phonons exhibiting $p$-, $d$- , $f$- , $g$- , $h$- , $i$- , $j$-wave PAM patterns, respectively. In contrast, if no symmetry operation in $G$ yields $\eta(\mathcal{T})\chi(\mathcal{R}) = -1$, then no such symmetry-enforced zero-PAM nodal surfaces exist. Under this condition, $j_{\textrm{m}}(\textbf{q})$ generically maintains a single sign in the BZ, thus corresponding to ferroaxial phonons with $s$-wave PAM pattern. 

A hallmark of altermagnets is the even-wave spin polarization in their band structures, which originates from the spin-only group symmetry $[\bar{C}_{2}|\mathcal{T}]$ within the spin-group framework and is independent of spatial inversion symmetry~\cite{smejkal_beyond_2022}. This even-wave feature can give rise to distinctive physical phenomena associated with altermagnets. In contrast, PAM in collinear magnets is governed by the MPG, which lacks the $[\bar{C}_{2}|\mathcal{T}]$ symmetry. This allows PAM to exhibit both even‑ and odd‑wave patterns. Specifically, nonzero PAM emerges only when inversion symmetry $\mathcal{P}$ and/or time‑reversal symmetry $\mathcal{T}$ (including the $\tau\mathcal{T}$ symmetry in antiferromagnets, where $\tau$ is a fractional translation) is broken: if $\mathcal{P}$ is preserved but $\mathcal{T}$ is broken, PAM exhibits an even‑wave pattern; if $\mathcal{T}$ (or $\tau\mathcal{T}$)  is preserved but $\mathcal{P}$ is broken, it displays an odd‑wave pattern; and if both symmetries are broken, PAM can adopt either an even‑ or an odd‑wave pattern. Based on these symmetry conditions and the resulting wave patterns, phonons in collinear magnets can be classified into four distinct types: (i) $\mathcal{PT}$-symmetric nonaxial phonons; (ii) Centrosymmetric (CS) even-wave axial phonons, which require breaking $\mathcal{T}$; (iii) Noncentrosymmetric (NCS) odd-wave axial phonons with no requirement on $\mathcal{T}$; (iv) NCS even-wave axial phonons with broken $\mathcal{T}$. Although types (ii) and (iv) both exhibit even-wave patterns, their physical origins differ significantly: the PAM in type (ii) originates purely from broken $\mathcal{T}$ symmetry, whereas in type (iv) it is contributed by both $\mathcal{T}$-breaking and $\mathcal{P}$-breaking terms. 

In collinear magnets, the phonon type is uniquely determined by its corresponding MPG. In Table I, we present the complete classification for all collinear MPGs, explicitly labeling the corresponding wave types. To be specific, $\mathcal{PT}$- and $\tau\mathcal{T}$-antiferromagnets host types (i) and (ii) phonons, respectively, while ferromagnets and altermagnets can harbor axial phonons belonging to any of the three types (ii) (iii) and (iv). Notably, alteraxial phonons featuring higher-order PAM wave patterns are not limited to altermagnets, but exist across all types of collinear magnets, spanning ferro-, antiferro-, and alter-magnets. Moreover, it is worth noting that the PAM wave pattern of alteraxial phonons in an altermagnet can differ from that of its electronic structure (e.g., in altermagnets hosting type (iii) phonons).\\



\begin{figure*}[t]
	\centering
	\includegraphics[width=0.95\textwidth]{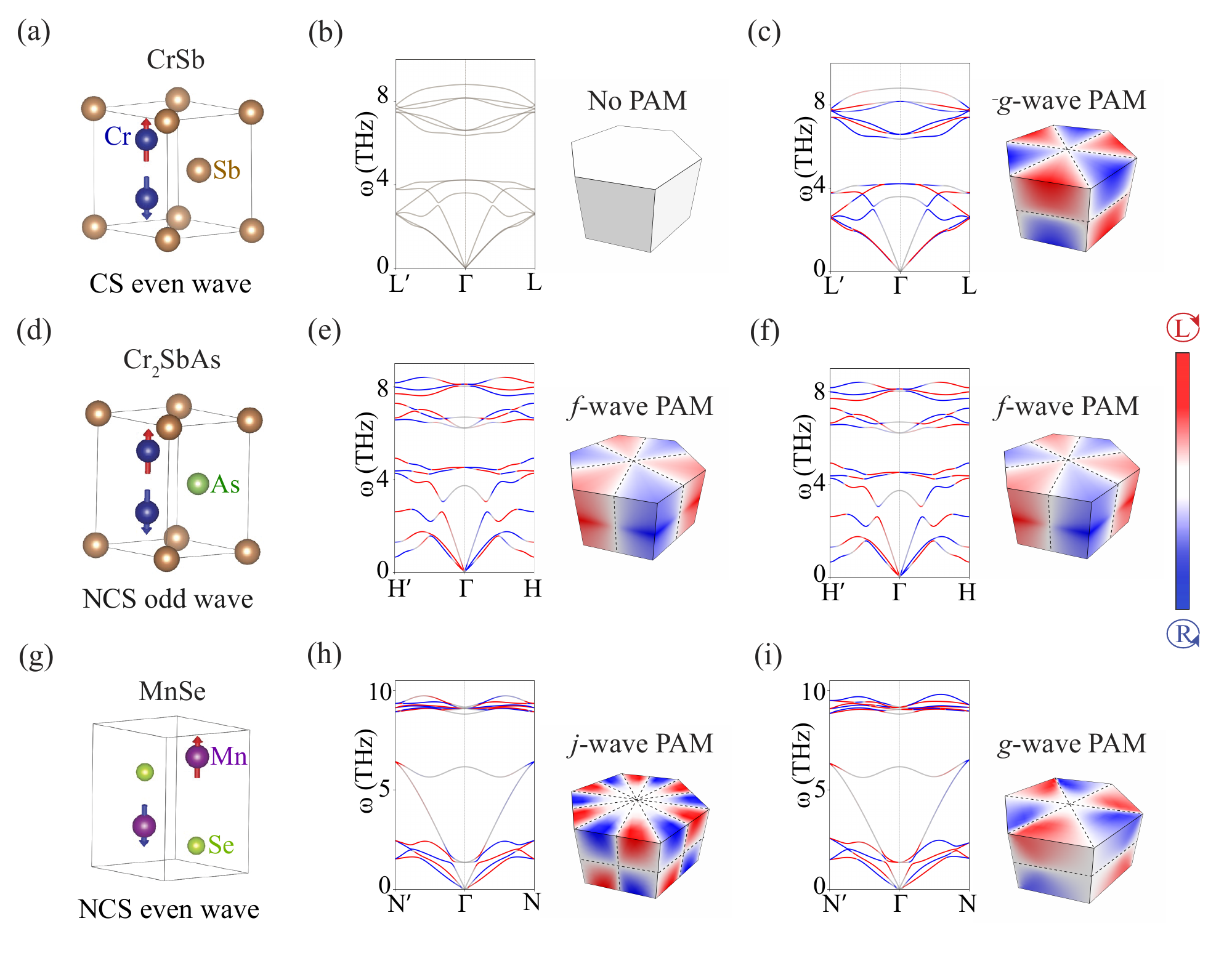}
	\caption{Phonon spectra and PAM patterns from \emph{ab initio} calculations on representative materials hosting three different types of alteraxial phonons. CrSb (CS even-wave, $g$-wave), Cr$_2$SbAs (NCS odd-wave, $f$-wave), and MnSe (NCS even-wave, $g$-wave) are selected as illustrative cases. For each material, the left column shows the crystal and magnetic structures, the middle (right) column presents the $\mathcal{T}$-symmetric ($\mathcal{T}$-broken) phonon spectra and PAM patterns around the $\Gamma$ point for the third phonon branch  in the absence (presence) of $\mathcal{T}$-breaking perturbations.}\label{fig3}
\end{figure*}

\noindent{\bf{High-throughput calculations.}}
To search for realistic collinear magnets hosting alteraxial phonons, we performed high-throughput calculations on magnetic materials from databases such as MAGNDATA~\cite{gallego_magndata_2016,gallego_magndata_2016-1} and Materials Project~\cite{jain_commentary_2013}. The calculation workflow, as depicted in Fig.~2(a), comprises two primary tasks after obtaining the crystal and magnetic structures, namely, determining the MPG-based phonon type and conducting \emph{ab initio} calcualtions of magnetic phonon spectra. Note that phonon spectra obtained from standard force-constant calculations are inherently  $\mathcal{T}$-symmetric [see the SM for details~\cite{supp}], and additional $\mathcal{T}$-breaking terms need to be accounted for to obtain magnetic phonon spectra. In magnetic materials, such $\mathcal{T}$-breaking terms typically arise from spin-phonon~\cite{lunghi_toward_2022,bonini_frequency_2023,kragskow_spinphonon_2023,ren_adiabatic_2024,wang_ab_2025} or magnon-phonon couplings~\cite{thingstad_chiral_2019,nomura_phonon_2019,wu_fluctuation-enhanced_2023,bao_direct_2023,wu_magnetic_2025,wang_magnetic_2024,ning_spontaneous_2024,weissenhofer_chiral_2025,bendin_d-wave_2025}, which usually constitute perturbative contributions to phonon spectra. In our calcualtions, the Raman-type spin-phonon coupling induced from electron-phonon interaction is incorporated as the $\mathcal{T}$-breaking perturbation~\cite{liu_model_2017}, which is implemented via a recently developed method by some of the authors~\cite{wang_ab_2025} (see the SM for details~\cite{supp}). 

In Figs. 2(b) and 2(c), we summarized our results for all 1029 collinear magnets from the MAGNDATA database [see SM~\cite{supp} for the results of Material Project database]. Of these 1,029 materials, 619 (60.2\%) host antiferro-nonaxial phonons. The remaining 410 materials host axial phonons, distributed as follows: 271 (26.3\%) are CS even-wave, 93 (9.0\%) are NCS odd-wave, and 46 (4.5\%) are NCS even-wave. Notably, we uncover 237 magnets that host alteraxial phonons consisting of both odd- and even-wave patterns, in addition to 173 materials exhibiting $s$-wave ferroaxial phonons.\\

\noindent{\bf{Representative materials.}} Inspired by the rapidly growing field of altermagnetism, we select three altermagnets as representative materials (see SM~\cite{supp} for other types of magnets), each hosting a distinct type of alteraxial phonon, CrSb~\cite{smejkal_beyond_2022} (CS even-wave), Cr$_2$SbAs (NCS odd-wave), and MnSe (NCS even-wave), as detailed below. 

CrSb exhibits an out-of-plane collinear magnetic order with antiparallel-aligned moments on Cr atoms [see Fig. 3(a)], which belongs to the MPG $6'/m'mm'$. Within the spin-group description, it is classified as a $g$-wave altermagnet \cite{smejkal_beyond_2022}. It possesses inversion symmetry with Cr atoms acting as the inversion center. Consequently, in the absence of $\mathcal{T}$-breaking perturbation, the phonon spetrum of CrSb preserves the combined $\mathcal{PT}$ symmetry, yielding nonaxial phonons with vanishing PAM [Fig. 3(d)]. When $\mathcal{T}$-breaking perturbation is incorparated, alteraxial phonons with $g$-wave PAM pattern emerge. This is evidenced by the phonon spectrum and PAM distribution [Fig. 3(g)] characterized by four nodal-surfaces with vanishing PAM crossing the $\Gamma$ point (three vertical surfaces connected by $C_{3z}$ and one horizontal $q_z=0$ surface enforced by $C_{2z}\mathcal{T}$ symmetry).

Cr$_2$SbAs is obtained by substituting one Sb site with As atom in CrSb [see Fig. 3(b)], thereby breaking the inversion symmetry and lowering the MPG to $\overline{6}'m'2$. Without the $\mathcal{T}$-breaking perturbation, the phonon spectrum exhibits $\mathcal{T}$-symmetry-enforced odd $f$-wave PAM pattern, with three vertical zero-PAM nodal surfaces passing through the $\Gamma$ point [Fig. 3(e)]. When the $\mathcal{T}$-breaking perturbation is considered, the $f$-wave pattern persists, as shown in Fig. 3(h). In fact, the $f$-wave pattern results from the removal of the $q_z=0$ nodal surface in CrSb due to the broken $C_{2z}\mathcal{T}$ symmetry. Interestingly, in striking contrast to its $f$-wave phonon spectrum, the electronic structure of Cr$_2$SbAs remains a $g$-wave altermagnet like CrSb.

MnSe is a noncentrosymmetric magnet hosting out-of-plane antiparallel alignment of Mn moments [see Fig.~3(c)], described by the MPG $6'mm'$. As shown in Fig.~3(f), the unperturbed phonon spectrum of MnSe with preserved $\mathcal{T}$ symmetry exhibits an odd $j$-wave PAM pattern, featuring seven zero-PAM nodal surfaces crossing the $\Gamma$ point, which are composed of six vertical surfaces enforced by six $C_{6z}$-connected $\sigma_{v}$ symmetries, and one horizontal surface ($q_z=0$) enforced by the combined $C_{2z}\mathcal{T}$ symmetry (see the SM for details~\cite{supp}). Under a $\mathcal{T}$-breaking perturbation, the $C_{2z}\mathcal{T}$ remains intact, whereas only a subset of three among the original six $\sigma_{v}$ symmetries survive. This shifts three vertical nodal surfaces from the $\Gamma$ point, thereby changing the $\Gamma$-centered PAM pattern from odd $j$-wave to even $g$-wave. It is worth noting that all NCS even-wave PAM patterns can be derived from a $\mathcal{T}$-preserving higher-order-wave odd-wave parent pattern with a higher symmetry, and the full derivation based on compatibility relations for each case is detailed in the SM.

{\bf{Conclusion.}} We have established a comprehensive group-theoretical framework to systematically classify axial phonons in collinear magnets and uncovered a complete catelogue of alteraxial phonons characterized by higher-order-wave PAM patterns. Our high-throughput screening identifies numerous candidate materials and \emph{ab initio} calculations are further performed on representative materials to confirm alteraxial phonons as well as their PAM patterns. Our work not only significantly enriches the family of axial phonons, but also opens a new avenue in phonon engineering via magneto-phononic effects.

{\bf{Acknowledgments.}} This work is partly supported by National Key Projects for Research and Development of China (Grants No. 2024YFA1409100, No. 2021YFA1400400, No. 2022YFA1403602, and No. 2023YFA1407001), the Natural Science Foundation of Jiangsu Province (Grants No. BK20253012, No. BK20252117, No. BK20233001, No. BK20243011, and No. BK20220032), the Natural Science Foundation of China (Grants No. 12534007, No. 92365203 and No.12174182) and the e-Science Center of Collaborative Innovation Center of Advanced Microstructures.

{\bf{Data availability.}} All data are available from the corresponding authors upon request.

{\bf{Code availability.}} The codes used in the study are available from the corresponding authors upon reasonable request.

{\bf{Competing financial interests.}} The authors declare no competing financial interests.

\bibliography{references}

\end{document}